\documentclass[12pt]{article}
\pdfoutput=1
\usepackage{latexsym, graphicx,cite} 
\usepackage{amsmath}
\usepackage{amssymb}
\usepackage[colorlinks=true, linkcolor=blue, bookmarks=true]{hyperref}

 \newcommand{\arXiv}[1]{\href{http://www.arXiv.org/abs/#1}{#1}}

\makeatletter
\renewcommand\section{\@startsection {section}{1}{\z@}%
                                   {-3.5ex \@plus -1ex \@minus -.2ex}
                                   {2.3ex \@plus.2ex}%
                                   {\normalfont\large\bfseries}}
\renewcommand\subsection{\@startsection{subsection}{2}{\z@}%
                                     {-3.25ex\@plus -1ex \@minus -.2ex}%
                                     {1.5ex \@plus .2ex}%
                                     {\normalfont\bfseries}}
\makeatother

\newcommand{\beq}{\begin{equation}}
\newcommand{\eeq}{\end{equation}}
\newcommand{\bea}{\beq}
\newcommand{\eea}{\eeq}
\newcommand{\dsty}{\displaystyle}
\newcommand{\eps}{\epsilon}
\newcommand{\om}{\omega}
\newcommand{\de}{\delta}
\newcommand{\del}{\partial}
\newcommand{\cP}{{\cal P}}
\newcommand{\s}{\sigma}

\begin{document}

\begin{titlepage}
\begin{flushright}
\phantom{arXiv:yymm.nnnn}
\end{flushright}
\vfill
\begin{center}
{\Large\bf AdS perturbations, isometries, selection rules\vspace{2mm}\\ and the Higgs oscillator}    \\
\vskip 15mm
{\large Oleg Evnin$^{a,b}$ and Rongvoram Nivesvivat$^a$}
\vskip 7mm
{\em $^a$ Department of Physics, Faculty of Science, Chulalongkorn University,\\
Thanon Phayathai, Pathumwan, Bangkok 10330, Thailand}
\vskip 3mm
{\em $^b$ Theoretische Natuurkunde, Vrije Universiteit Brussel and\\
The International Solvay Institutes\\ Pleinlaan 2, B-1050 Brussels, Belgium}

\vskip 3mm
{\small\noindent  {\tt oleg.evnin@gmail.com, rongvoramnivesvivat@gmail.com}}
\vskip 10mm
\end{center}
\vfill

\begin{center}
{\bf ABSTRACT}\vspace{3mm}
\end{center}

Dynamics of small-amplitude perturbations in the global anti-de Sitter (AdS) spacetime is restricted by selection rules that forbid effective energy transfer between certain sets of normal modes. The selection rules arise algebraically because some integrals of products of AdS mode functions vanish. Here, we reveal the relation of these selection rules to AdS isometries. The formulation we discover through this systematic approach is both simpler and stronger than what has been reported previously. In addition to the selection rule considerations, we develop a number of useful representations for the global AdS mode functions, with connections to algebraic structures of the Higgs oscillator, a superintegrable system describing a particle on a sphere in an inverse cosine-squared potential, where the AdS isometries play the role of a spectrum-generating algebra. 

\vfill

\end{titlepage}


\section{Introduction}

The notion that symmetries and mode multiplet structures of linearized perturbations constrain possible effects of nonlinearities in the weakly nonlinear (small amplitude) regime is familiar from elementary mechanical settings, such as nonlinear vibrations of crystalline lattices, see e.g. \cite{chechin}. In this article, we shall deal with the implications of this phenomenon for weakly nonlinear relativistic fields in anti-de Sitter (AdS) geometry, a maximally symmetric spacetime that plays a central role in the AdS/CFT correspondence. We shall also pay special attention to the connections between the algebraic structures arising in this context and algebras of conserved quantities of the Higgs oscillator,\footnote{No connection to the Higgs boson, or the Brout-Englert-Higgs mechanism.} a well-known superintegrable mechanical system describing a particle on a sphere in a central potential varying as the inverse cosine-squared of the polar angle \cite{Higgs,Leemon}.

Dynamics of small amplitude perturbations in AdS backgrounds has attracted a considerable amount of attention since the pioneering numerical observations of \cite{BR}. Computer simulations indicate that certain initial data of amplitude $\eps$ collapse to form black holes on time scales of order $1/\eps^2$, no matter how small the amplitude is. Attempts to analyze this problem using naive perturbative expansions in powers of the amplitude $\eps$ are plagued by secular terms which grow in time and invalidate the expansion precisely at time scales of physical interest. Resummed (improved) expansions can be constructed, with the formalism featuring effective flow equations describing slow energy transfer between linearized normal modes due to nonlinearities. These flow equations can be shown to accurately describe the dynamics on time scales of order $1/\eps^2$, which are precisely the time scales of interest. Further review of these approaches with references to extensive original literature can be found in \cite{MRlectures, BizonReview, CEproc}.

Once the resummation procedure we described above has been applied, the flow equations can be derived and they contain a number of terms, each corresponding to energy transfer between a certain set of modes. The coefficients of these terms (the ``interaction coefficients'') are certain integrals of products of the corresponding mode functions (of linearized fields), whose precise structure depends on the form of nonlinearities in the original equations of motion. It turns out that a large fraction of these interaction coefficients vanishes due to special properties of the AdS mode functions \cite{CEV1}. These {\em selection rules} have a number of immediate consequences for the dynamics of the effective flow equations. For example, they enhance the submanifolds of special solutions in which effective energy transfer between the modes does {\em not} occur (the so-called quasiperiodic solutions) \cite{CEV1}, as well as the set of conserved quantities \cite{CEV2} of the flow equations. The enhanced set of conserved quantities enforces ``dual cascades,'' meaning that any energy flow to shorter wave-lengths (a necessary precursor of black hole formation in our weak field setting) has to be accompanied by some energy flowing to longer wave-lengths \cite{BGLL}.

The selection rules arise algebraically because certain integrals of products of the AdS mode functions vanish. This is proved in practice by considering explicit expressions for the mode functions in terms of Jacobi polynomials, and then using arguments based on orthogonality properties of the Jacobi polynomials. For the case of spherically symmetric perturbations of a fully dynamical asymptotically AdS geometry coupled to a scalar field, this sort of proof for the selection rules has been given in \cite{CEV1}. It is often beneficial to consider a toy model in which gravitational interactions are turned off and one is dealing with the dynamics of a self-interacting probe scalar field in a fixed AdS background \cite{BKS1,BKS2}. This system is much simpler and allows analysis of perturbation theory without assuming spherical symmetry,\footnote{The non-spherically-symmetric case with full-fledged gravitational interactions is forbiddingly complicated, though some limited amount of progress has been made \cite{DHS,HS}.} including a large and powerful set of selection rules \cite{Yang} (a compact proof of selection rules for the same system with spherical symmetry imposed can be found in footnote 3 of \cite{CEV2}).

At a practical level, one could survive with brute force derivations of the selection rules using properties of the Jacobi polynomials. Such derivations, however, make the qualitative origin of the selection rules very intransparent. It is natural to believe that the selection rules are mandated by rich symmetries of the underlying AdS background, which is a maximally symmetric spacetime with an isometry group $SO(d,2)$ for AdS$_{d+1}$. Such suspicions have been voiced already in \cite{CEV1} and \cite{Yang}. To give a simple example of how symmetries may enforce selection rules, one might consider an integral of a product of spherical harmonics of the form $\int  Y_{l_1m_1}\cdots Y_{l_Nm_N}d\Omega$. One can show that this integral vanishes unless the identity representation of the rotational group is contained in the direct product of the representations corresponding to each of the spherical harmonics. This forces each $l_i$ to be less than or equal to the sum of all the other $l_i$.

Application of symmetries to selection rules for integrals of products of the AdS mode functions is less straightforward than the above example involving spherical harmonics. One first has to decide what is the symmetry defining the multiplets of AdS mode functions of a given frequency. This cannot be the AdS isometry group, since the boost generators change the frequency. Furthermore, the mode functions form large multiplets, implying that the relevant symmetry group is bigger than, say, the obvious group of spatial rotations.

In \cite{EK}, it was shown that mode functions of the same frequency in global AdS$_{d+1}$ form multiplets of a hidden $SU(d)$ symmetry (including the $SO(d)$ spatial rotations). This is demonstrated by relating the mode function equation to the Schr\"odinger equation of the Higgs oscillator, which is known to possess a hidden $SU(d)$ symmetry \cite{Higgs}. Historically, the relevant energy eigenvalue problem was solved in \cite{lakshesw} from a purely quantum-mechanical perspective. The observed large level degeneracy prompted an investigation into enhanced symmetries, which resulted in the discovery of the hidden $SU(d)$ symmetry \cite{Higgs} and subsequent extensive studies of the related algebraic structures in the mathematical quantum mechanics community. On the AdS side, explicit expressions for the mode functions in terms of Jacobi polynomials can be found, e.g., in \cite{BF,HKLL,FKPS}. We see no evidence of connections between the two bodies of literature treating these closely related structures in two separate guises (quantum mechanics of the Higgs oscillator and field dynamics in AdS spacetimes), until a link was made in \cite{EK}.

Even though the hidden $SU(d)$ group clearly explains the multiplet structure of global AdS$_{d+1}$ mode functions involved in the selection rules, it is difficult to employ for proving the selection rules directly, at least with the current state of knowledge \cite{EK}. The reason is that no explicit construction of the $SU(d)$ generators exists due to the nonlinear nature of the Higgs oscillator and difficulties in resolving the ordering ambiguities while quantizing the classical generators \cite{Higgs,Leemon,interbasis}. In this paper, we take a different approach and develop representations for global AdS mode functions in which the isometries of AdS act manifestly as a spectrum-generating algebra. This allows one to write explicit formulas for the global AdS mode functions in terms of isometry-based raising operators acting on the mode function of the lowest frequency (incidentally, this representation makes the $SU(d)$ symmetric tensor nature of the mode function multiplets completely manifest). Using such formulas, we succeed in proving the selection rules entirely in terms of AdS isometries. The result is both simpler and stronger than what has been reported in \cite{Yang}. In \cite{Yang}, only the radial dependences of mode functions were considered, and a selection rule was proved for the case when the number of mode functions inside the integral and the number of spatial dimensions of AdS are not both odd. This restriction is unnecessary, however, since  in the remaining case when they are both odd, the angular integral (not considered in \cite{Yang}) vanishes automatically. Our derivations do not separate the mode functions into radial and angular parts and produce the stronger version of the selection rules (applying to any number of mode functions inside the integral, and any dimension of the AdS) in a straightforward manner.

The techniques we employ mostly revolve around the use of a flat embedding space, in which the AdS can be realized as a hyperboloid (our approach is largely consonant with the presentation of \cite{kaplan}). This allows a simple explicit treatment of the AdS isometries, and a simple characterization of the AdS mode functions (or the Higgs oscillator energy eigenstates) in terms of homogeneous polynomials of the flat space Cartesian coordinates. Such treatment is parallel to the usual relation between spherical harmonics on a $d$-sphere and the solid harmonics, which are homogeneous polynomials in a $(d+1)$-dimensional flat space. A number of representations we derive should be of interest from a purely quantum-mechanical Higgs oscillator perspective. Thus, the isometries of AdS provide an $so(d,2)$ spectrum-generating algebra for the Higgs oscillator, the conserved rank two symmetric tensor emerges from quadratic combinations of the elements of this $so(d,2)$ algebra, the energy eigenstates of the Higgs oscillator in $d$ dimensions are realized as a peculiar subsector of harmonic oscillator motion in a flat pseudo-Euclidean $(d+2)$-dimensional space, and the obscure `gnomonic' coordinates on the sphere \cite{Higgs}, in which the conserved quantities take a simple form, arise with inevitability from the Cartesian coordinates on this $(d+2)$-dimensional space. This higher-dimensional geometrization of the sophisticated algebraic structures enjoyed by the Higgs oscillator is both highly visual and potentially useful.

The paper is organized as follows: In section 2, we review the basics of constructing the effective flow equations for small amplitude perturbations in AdS, and the emergence of selection rules. Section 3 is a brief summary of \cite{EK}, where the structure of mode function multiplets is explained and related to the Higgs oscillator problem. Section 4 develops a representation for the global AdS mode functions in terms of homogeneous polynomials in the flat embedding space and discusses how the AdS isometries operate in this context. Section 5 derives selection rules using a part of the isometry group as raising and lowering operators for the mode functions. Readers familiar with the context of nonlinear perturbation theory in AdS and specifically interested in the selection rule problem may read this section largely independently, though the explicit representations developed in section 3 make the isometries action much more concrete. Section 6 discusses the spectrum-generating algebra for the Higgs oscillator provided by the AdS isometries and the construction of conserved quantities as quadratic combinations of the isometry generators.


\section{Nonlinear AdS perturbations and selection rules}

We very briefly review the basics of nonlinear perturbation theory in AdS, referring the reader to \cite{CEproc} and the original publications cited therein for further details.

The metric of AdS$_{d+1}$ (with the AdS radius set to 1 by rescaling) is given by
\beq
ds^{2} = \frac1{\cos^{2}x}\left(-dt^{2} + dx^{2} + \sin^{2}x\, d\Omega_{d-1}^{2}\right),
\label{AdSmetric}
\eeq
where $d\Omega_{d-1}^{2}$ is the line element of an ordinary $(d-1)$-sphere parametrized by angles collectively denoted as $\Omega$. The bulk of the research efforts (see \cite{MRlectures, BizonReview, CEproc} and references therein) has been directed at studying spherically symmetric perturbations of the above metric (perturbations only depending on $t$ and $x$) coupled to a scalar field. It is instructive, however, to examine (following \cite{BKS1,BKS2,Yang}) a simpler `toy model' in which a self-interacting scalar field evolves in a frozen geometry of the form (\ref{AdSmetric}). The action for the field is taken to be
\beq
S = \int d^{d+1}x \sqrt{-g} \, \left(\dfrac{1}{2}g^{\mu\nu}\partial_{\mu}\phi \ \partial_{\nu}\phi + \dfrac{m^2}{2}\phi^2+ \dfrac{\phi^{N+1}}{(N+1)!} \right),
\eeq
resulting in equations of motion of the form
\beq\label{scalareom}
 \cos^{2}x\ \left(
-\partial_t^2\phi + \frac{1}{\tan^{d-1} x} \partial_x(\tan^{d-1} x \partial_x \phi)+\frac{1}{\sin^2x} \Delta_{\Omega_{d-1}}\phi \right)
-m^2 \phi  = \dfrac{\phi^{N}}{N!} ,
\eeq
where $\Delta_{\Omega_{d-1}}$ is the ordinary $(d-1)$-sphere Laplacian. The scalar field `toy model' of this sort encapsulates many of the interesting properties of nonlinear perturbation theory of the full gravitational case, including the selection rules, and permits analyzing them with greater ease, including completely general non-spherically-symmetric perturbations.

If one is interested in the weakly nonlinear (small amplitude) regime, one has to start by solving the linearized system in which the right-hand side of (\ref{scalareom}) is neglected. Separation of variables leads to solutions of the form
\beq
\label{GlobalModes}
\phi_{\mbox{\scriptsize linear}}(t,x,\Omega) = \sum_{n = 0}^\infty \sum_{l,k} (A_{nlk}\ e^{i\omega_{nlk}t}+\bar A_{nlk}\ e^{-i\omega_{nlk}t}) e_{nlk}(x, \Omega),
\eeq
where $A_{nlk}$ are arbitrary complex amplitudes and
\beq
\omega_{nlk}=\delta+2n + l, \label{omega}
\eeq
with
$\delta= \frac{d}{2} + \sqrt{{d^2 \over 4} + m^2}$. The mode functions can be read off \cite{BF,HKLL,FKPS} and are given by
\bea
e_{nlk}(x, \Omega)=\cos^\delta \! x \ \sin^l \! x\ P_n^{\left(\delta - \frac{d}{2},\,l + \frac{d}{2} - 1\right)}
(-\cos 2 x) \ Y_{lk}(\Omega).
\label{modefunctions}
\eea
$Y_{lk}$ are spherical harmonics in $(d-1)$ dimensions, with $l(l+d-2)$ being the eigenvalue of the corresponding sphere Laplacian, and $k$ labelling all the different harmonics contained in a given $l$-multiplet. $P^{(a,b)}_n(y)$ are Jacobi polynomials orthogonal with respect to the measure $(1-x)^a(1+x)^b$ on the interval $(-1,1)$. We shall not be careful about mode function normalizations below, since our objective is to prove that some of their integrals are exactly zero. All the equations are meant to hold up to the mode function normalization. The mode functions satisfy the following equation
\bea
\left(\frac{1}{\tan^{d-1} x} \partial_x(\tan^{d-1} x \partial_x )+\frac{1}{\sin^2x} \Delta_{\Omega_{d-1}}-\frac{m^2}{\cos^2 x}\right)e_{nlk}(x, \Omega)=-\omega_{nlk}^2 \ e_{nlk}(x, \Omega),
\label{adsmode}
\eea

Armed with the linearized solutions (\ref{GlobalModes}), one could try to analyze the leading nonlinear corrections by performing a weak field expansion of the form
\beq
\phi=\eps \phi_{\mbox{\scriptsize linear}} + \eps^N  \phi_{\mbox{\scriptsize corr.}} +\cdots
\label{naive}
\eeq
This approach is plagued by secular terms in $\phi_{\mbox{\scriptsize corr.}}$ which grow in time and invalidate the expansion precisely where one is trying to obtain predictions of qualitative relevance. The naive perturbative expansion can be resummed with a variety of methods, leading to flow equations describing slow energy transfer between the normal modes. The quickest (and equivalent) way to present these flow equations, however, is not to start with the naive perturbative expansion (\ref{naive}), but rather to employ a procedure known as time-averaging.

The first step of time-averaging is to switch to the `interaction picture' (`periodic standard form' in the mathematical parlance) in (\ref{scalareom}). One first expands the exact $\phi$ in linearized normal modes
\bea
\label{phiModes}
 \phi (t,x,\Omega) = \sum_{nlk} c_{nlk}(t) e_{nlk}(x, \Omega),
\eea
and then introduces complex amplitudes $\alpha(t)$ that would have been constant had the self-interactions been turned off:
\beq
c_{nlk}=\eps\left(\alpha_{nlk}e^{i\om_{nlk} t}+\bar \alpha_{nlk}e^{-i\om_{nlk} t}\right),\qquad \dot c_{nlk}=i\eps\,\om_{nlk}\left(\alpha_{nlk}e^{i\om_{nlk} t}-\bar \alpha_{nlk}e^{-i\om_{nlk} t}\right).
\eeq
This leads to the equation
\begin{align}\label{interaction}
2i\om_{nkl}&\dot\alpha_{nkl} = \dfrac{\eps^N e^{-i\omega_{nlk}t}}{N!}\sum\limits_{n_1l_1k_1}\cdots\sum\limits_{n_Nl_Nk_N} C_{nlk|n_1l_1k_1|\cdots|n_{N}l_{N}k_{N}}\\
\nonumber&\times(\alpha_{n_1l_1k_1}\ e^{i\omega_{n_1l_1k_1}t}+\bar \alpha_{n_1l_1k_1}\ e^{-i\omega_{n_1l_1k_1}t})\cdots(\alpha_{n_{N}l_{N}k_{N}}\ e^{i\omega_{n_{N}l_{N}k_{N}}t}+\bar \alpha_{n_{N}l_{N}k_{N}}\ e^{-i\omega_{n_{N}l_{N}k_{N}}t}).
\end{align}
Note that $\om_{nlk}$ do not depend on $k$, but we keep the $k$ index to be able to track later which mode the frequency is referring to. The interaction coefficients $C$ are given by
\bea
C_{nlk|n_1l_1k_1|\cdots|n_{N}l_{N}k_{N}}=\int dx d\Omega\, \frac{\tan^{d-1}x} {\cos^2x}\,e_{nlk} e_{n_1l_1k_1}\cdots e_{n_Nl_Nk_N}.
\label{Ccoeff}
\eea
While $\alpha_{nlk}$ governed by (\ref{interaction}) vary on time scales of order $1/\eps^N$, the right-hand side of (\ref{interaction}) contains terms oscillating on time scales of order $1$. The main point of the averaging method is that the effect of these terms `averages out' and they can be simply discarded. This can be embodied in mathematical theorems stating that the resulting approximation is accurate on time scales of order $1/\eps^N$.

Non-oscillating terms on the right-hand side of (\ref{interaction})  come from `resonant' sets of frequencies satisfying
\beq
\om_{nlk}=\pm \omega_{n_1l_1k_1}\pm\omega_{n_2l_2k_2}\pm\cdots\pm \omega_{n_Nl_Nk_N},
\label{res}
\eeq
where all the plus-minus signs are independent. With only these terms retained, the resulting `flow equation' takes the following schematic form:
\beq
2i\om_{nkl}\dot\alpha_{nkl} = \dfrac{\eps^N}{N!}\sum\limits_{\mbox{\scriptsize resonance}} C_{nlk|n_1l_1k_1|\cdots|n_{N}l_{N}k_{N}} \overset{?}{\alpha}_{n_1l_1k_1}\cdots \overset{?}{\alpha}_{n_{N}l_{N}k_{N}},\vspace{-2.5mm}
\label{flow}
\eeq
where $ \overset{?}{\alpha}$ denotes either $\alpha$ or $\bar\alpha$, depending on whether the corresponding $\omega$ appears with plus or minus sign in the resonance relation (\ref{res}) corresponding to the particular term in question.

For the case of a massless field, $\delta=d$ and all the frequencies (\ref{omega}) are integer. This gives the resonance condition (\ref{res}) a tremendous number of solutions. However, explicit computations show that the interaction coefficients defined by (\ref{Ccoeff}) corresponding to some choices of signs in (\ref{res}) vanish. In particular, the $C$ coefficients corresponding to all plus signs in (\ref{res}) always vanish. The subject of dynamical consequences of this sort of {\em selection rules} has been discussed at length in the literature, and we shall not delve into that here.   In the rest of the paper, giving a maximally transparent geometrical understanding to selection rules in the interaction coefficients defined by (\ref{Ccoeff}) will form one of our main objectives.


\section{AdS mode function multiplets and the Higgs oscillator}

Before we proceed with investigating the role of symmetries in the selection rules of nonlinear perturbation theory, where they restrict (\ref{Ccoeff}), we note that some intriguing symmetry related patterns are seen in AdS already at the linearized level. Indeed, the multiplets of mode functions $e_{nlk}$  with a given frequency $\om_{nlk}=\de+2n+l$ are abnormally large. Not only does the frequency not depend on $k$ within each given $l$-multiplet (this is trivially dictated by the rotation symmetry of the $(d-1)$-sphere in the AdS metric), but it also depends on $n$ and $l$ in one particular combination $2n+l$ which leads to many distinct values of $n$ and $l$ producing the same frequency.  Such high degeneracy is familiar from quantum-mechanical examples, such as the hydrogen atom, where its existence is explained by the presence of hidden symmetries.

The issue of the symmetry origins of the abnormally high degeneracies in (\ref{omega}-\ref{modefunctions}) has been resolved in \cite{EK}. One simply rewrites (\ref{adsmode}) in terms of $\tilde e_{nlk} \equiv e_{nlk}/\cos^{(d-1)/2} x$, obtaining
\beq
\left(-\Delta_{\Omega_d}+V(x)\right) \tilde e_{nlk}=E_{nlk} \tilde e_{nlk}, \label{schrod}
\eeq
with
\beq
V(x)=\frac{(2\de-d)^2-1}{4\cos^2 x}\qquad\mbox{and}\qquad E_{nlk}=\omega_{nlk}^2-\frac{(d-1)^2}4.
\label{hggs}
\eeq
The $d$-sphere Laplacian in the above expression is given by the standard formula:
\beq 
\Delta_{\Omega_d} \equiv  \frac{1}{\sin^{d-1} x} \partial_x(\sin^{d-1} x \ \partial_x)+\frac{1}{\sin^2x} \Delta_{\Omega_{d-1}}. 
\label{dsphL}
\eeq
Geometrically, the variable redefinition we have performed is linked to the conformal transformation that maps AdS to one half of the Einstein static universe. Note that even though the Laplacian is defined on the whole sphere, in practice only one half of the sphere with $x \in [0,\pi/2)$ plays a role because of the infinite `potential well' provided by $V(x)\sim 1/{\cos^2 x}$.

One can understand (\ref{schrod}) as the Schr\"odinger equation for a particle on a $d$-sphere moving in a potential proportional to $1/{\cos^2 x}$. The hidden symmetries of this system, which is often called the Higgs oscillator, have been widely discussed following the investigations of \cite{Higgs,Leemon}. It is known that (\ref{schrod}) admits a hidden $SU(d)$ symmetry group which includes explicit $SO(d)$ rotations around the point $x=0$ as its subgroup. The mode functions of the $N$th energy level above the ground state transform in the fully symmetric rank $N$ tensor representation of the $SU(d)$. The generators of the $SU(d)$ are not known explicitly \cite{Higgs,Leemon,interbasis}. Even though the conserved quantites of the classical problem corresponding to the Schr\"odinger equation (\ref{schrod}) are known explicitly and can be assembled in combinations that form an $su(d)$ algebra with respect to taking Poisson brackets \cite{Higgs}, it is not known how to resolve the ordering ambiguities while quantizing these classical conserved quantities in a way that reproduces the Lie algebra at the quantum level, except for the relatively simple $d=2$ case \cite{Higgs}.

Implications of the hidden $SU(d)$ symmetry for (\ref{schrod}) and the AdS mode functions are discussed in greater detail in \cite{EK}, and we shall limit ourselves here to the multiplicity counting and some basic comments. As we have remarked, the $N$th energy level of the Higgs oscillator supports the fully symmetric rank $N$ tensor representation of the $SU(d)$ group. The degeneracy is then just the dimension of this representation, which can be counted as the number of possible sets of $d$ non-negative integers $p_i$ satisfying $\sum_i p_i=N$ (the integers $p_i$ simply encode how many times the index value $i$ occurs among the indices of the fully symmetric tensor). The number of partitions of $N$ into $p_i$ is easily counted as the number of ways to place $d-1$ separators in $N+d-1$ positions (with the numbers of empty places left between the $d-1$ separators interpreted as $p_1$, $p_2$, ... $p_d$). Thus the degeneracy of level $N$ in $d$ dimensions is just
\beq
\#(N,d)=\frac{(N+d-1)!}{N!(d-1)!}.
\label{degenr}
\eeq
On the other hand, we have an explicit specification of the spectrum (\ref{omega}), which says that level $N$ consists of multiplets with angular momentum $N$, $N-2$, $N-4$,..., one copy each. Thus, it should be possible to recover the degeneracy (\ref{degenr}) by summing the dimensions of the above angular momentum multiplets. The angular momentum multiplets are simply fully symmetric {\em traceless} tensors of the corresponding rank. The number of fully symmetric traceless tensors of rank $N$ can be obtained as the number of fully symmetric tensors of rank $N$ minus the number of fully symmetric tensors of rank $N-2$ (which are just the traces that need to be subtracted). Thus, we get for the multiplicity $(\#(N,d)-\#(N-2,d))+(\#(N-2,d)-\#(N-4,d))+\cdots=\#(N,d)$, recovering (\ref{degenr}). Explicit decompositions of $SU(d)$ representations into rotational $SO(d)$ multiplets can be found in \cite{hamermesh} and conform to the notion of (\ref{modefunctions}) with a given value of frequency forming appropriate $SU(d)$ multiplets. We note that the multiplet structures of the Higgs oscillator are identical to those of the ordinary (flat space) isotropic harmonic oscillator, whose hidden $SU(d)$ symmetry can be made manifest with ease.


\section{Mode functions in the embedding space}

As mentioned in \cite{EK} the lack of explicitly constructed generators for the Higgs oscillator's hidden symmetries makes it difficult to investigate the consequences of these symmetries in integral expressions like (\ref{Ccoeff}). We shall therefore approach the problem from a different perspective trying to characterize more transparently the spaces spanned by mode functions (\ref{modefunctions}).

AdS$_{d+1}$ can be realized as a hyperboloid in a flat pseudo-Euclidean space of dimension $d+2$ with the metric $\eta_{IJ}=\mbox{diag}(-1,-1,1,\cdots,1)$, being defined by the equation
\beq
\eta_{IJ}X^IX^J\equiv -X^2-Y^2+X^iX^i=-1, \qquad I,J\in \left\{X,Y,1,\cdots,d\right\}.
\label{embedding}
\eeq
It is known that many properties of the AdS spacetime become more transparent if viewed from the embedding space. Thus, the isometries of AdS are the obvious linear transformations of $X^I$ leaving $\eta_{IJ}$ invariant, which coincides with the definition of $SO(d,2)$.

Not only the isometries, but also properties of geodesics in AdS become extremely transparent if viewed from the embedding space.  Each AdS geodesic lies in a 2-plane in the embedding space passing through the origin. AdS has the remarkable property that all geodesics are closed, which is something of a `classical' precursor to the peculiar Klein-Gordon frequency spectrum (\ref{omega}) and simple periodicity properties of solutions to the Klein-Gordon equation. (Spaces in which all geodesics are closed are known to possess very special features \cite{besse}. For example, the set of all geodesics of such spaces is itself a manifold. This space,  for AdS, is a pseudo-Euclidean version of a Grassmanian.) Viewed from the embedding space, the geodesic motion on AdS$_{d+1}$ becomes a subset of trajectories of an ordinary harmonic oscillator in the $(d+2)$-dimensional embedding space (see, e.g., \cite{kaplan}). On the other hand, the shapes of geodesics in the coordinates (\ref{AdSmetric}) can be straightforwardly related to the orbits of the Higgs oscillator, as we demonstrate in appendix A.

Given the dramatic simplification of the geodesic motion if viewed from the embedding space, it is natural to look for a similar simplification in solutions to the Klein-Gordon equation and the corresponding mode functions (\ref{modefunctions}). In particular, one could try to find a relation to quantum harmonic oscillator motion in the pseudo-Euclidean embedding space. We shall see below that such a picture can indeed be developed. Our strategy is very similar to the embedding of a sphere in an ordinary Euclidean space and extending the spherical harmonics into solid harmonics $r^l Y_{lm}(\Omega)$. Solid harmonics are simply homogeneous polynomials satisfying the Laplace equation and their properties are much more transparent than those of spherical harmonics viewed on the sphere. The pseudo-Euclidean version of the same story, leading to a construction of the mode functions (\ref{modefunctions}) in terms of homogeneous polynomials, is similar, even if less straightforward.

Mode functions (\ref{modefunctions}), which are functions of $x$ and $\Omega$ can be trivially extended to the entire hyperboloid (\ref{AdSmetric}) as $e^{i\om_{nlk}t} e_{nlk}(x,\Omega)$. We then have to decide how to conveniently extend this expression into the embedding space. This is essentially done by trial-and-error, but the simple prototype treatment of spherical harmonics provides some hints. Just like the spherical harmonics are extended in the Euclidean space by foliating the Euclidean space with spheres and parametrizing it with the standard coordinates on each sphere plus the sphere radius, one can foliate the flat target space with hyperboloids $\eta_{IJ}X^IX^J=-L^2$ and parametrize it with the coordinates of the form (\ref{AdSmetric}) on each hyperboloid, together with $L$ (this foliation does not cover the entire embedding space, but our derivations will not be impeded by this flaw). More specifically, we parametrize (a part of) the embedding space (including the AdS hyperboloid)  by $(L,t,x,\Omega)$ as
\beq
X=\frac{L\cos t}{\cos x},\qquad Y=\frac{L\sin t}{\cos x},\qquad X^i=Ln^i(\Omega)\tan x,
\label{embed}
\eeq
where $n^i(\Omega)$ is the unit vector pointing in the direction given by $\Omega$. $L=1$ corresponds to our original hyperboloid (\ref{AdSmetric}). It is more convenient to view the same foliation after introducing polar coordinates $(S,T)$ in the $(X,Y)$-plane and spherical coordinates $(R,\Omega)$ instead of $X^i$, so that the embedding space metric is
\beq
ds^2_{\mbox{\scriptsize target}}\equiv \eta_{IJ}dX^IdX^J= -dS^2-S^2dT^2+dR^2+R^2d\Omega^2_{d-1},
\label{targetsphere}
\eeq
and (\ref{embed}) takes the form
\beq
S=\frac{L}{\cos x},\qquad T=t,\qquad R=L\tan x,
\eeq
with the $\Omega$-coordinates identical on the target space and on the foliating hyperboloids. The $x$-coordinate of our hyperbolic foliation can then be expressed through the coordinates on the target space as $x=\arcsin(R/S)$. The idea is then to extend functions $\phi(t,x,\Omega)$ defined on the AdS hyperboloid in the embedding space as  $\phi(T,\arcsin(R/S),\Omega)$, possibly multiplied with a suitably chosen function of $L\equiv\sqrt{S^2-R^2}$. In other words, the numerical data are copied from the AdS hyperboloid to other hyperboloids of the foliation (\ref{embed}) and then possibly scaled independently (but uniformly) on each hyperboloid as a function of its radius.

We can now apply the type of embedding space extensions we have outlined above to $\phi(t,x,\Omega)$ satisfying the AdS Klein-Gordon equation,
\beq
\label{AdSwaveeq}
\Big(\Box_{AdS}-m^2\Big)\phi=0.
\eeq
It is a matter of straightforward algebra to show that
\beq
\Phi(T,S,R,\Omega)\equiv e^{(S^2-R^2)/2}(S^2-R^2)^{-\delta/2}\phi\left(T,\arcsin\frac{R}{S},\Omega\right)
\label{targetsubst}
\eeq
satisfies the Schr\"{o}dinger equation for a harmonic oscillator on the pseudo-Euclidean space (\ref{targetsphere}):
\beq
\label{Schreq}
\frac12\Big(\Delta_{\mbox{\scriptsize target}}+(S^2-R^2)\Big)\Phi=\left(\delta-\frac{d}2-1\right)\Phi,
\eeq
where
\beq
\label{targetop}
\Delta_{\mbox{\scriptsize target}} =-\frac{1}{S}\frac{\partial}{\partial S}\left(S\frac{\partial}{\partial S}\right)-\frac{1}{S^2}\frac{\partial^2}{\partial T^2}+\frac{1}{R^{d-1}}\frac{\partial}{\partial R}\left(R^{d-1}\frac{\partial}{\partial R}\right)+\frac{1}{R^2}\Delta_{\Omega_{d-1}}.
\eeq
Equation (\ref{Schreq}) validates our expectation that the AdS Klein-Gordon equation can be related to quantum harmonic oscillator motion in the embedding space.

One can take a particular solution of the Klein-Gordon equation, based on one normal mode, $\phi=e^{i\om_{nlk} t} e_{nlk}(x,\Omega)$ and extend it in the target space using the above procedure (we shall omit the indices of $\om$ from now on for brevity, keeping in mind that $\om=\de+2n+l$). This yields
\beq
{\cal E}_{nlk}=\frac{e^{i\om T}}{S^{\om}} 
\left(S^{2n} P^{(l+\frac{d}{2}-1,\de-\frac{d}{2})}_n
\left(1-\frac{2R^2}{S^2}\right)\right)
\left(R^l Y_{lk}(\Omega)\right)e^{(S^2-R^2)/2}.
\label{Etarget}
\eeq
Note that the first brackets contain a homogeneous polynomial of degree $n$ in $S^2$ and $R^2$, while the second brackets contain a homogeneous polynomial of degree $l$ in $X^i$ (the latter are just the solid harmonics).

It turns out that the subspace of solutions to (\ref{Schreq}) spanned by functions of the form (\ref{Etarget}) is straightforwardly characterized in the Cartesian coordinates. Namely, any function of the form
\beq
\frac{1}{(X-iY)^\omega} {\cal P}(X^2+Y^2,X^i)\, e^{(X^2+Y^2-X^iX^i)/2},
\label{cart}
\eeq
satisfying the Schr\"odinger equation (\ref{Schreq}), where $\cal P$ is a homogeneous polynomial of degree $\omega-\de$, ${\cal P}((\lambda X)^2+(\lambda Y)^2, \lambda X^i) =\lambda^{\omega-\de} {\cal P}(X^2+Y^2,X^i)$, can be expanded in terms of (\ref{Etarget}). To see this, one can always re-express (\ref{cart}) in terms of the polar-spherical coordinates (\ref{targetsphere}) and expand it in spherical harmonics as
\beq
\frac{e^{i\om T}}{S^{\om}} e^{(S^2-R^2)/2}\sum_{lk} {\cal P}_{lk}(S,R) Y_{lk}(\Omega).
\eeq
${\cal P}_{lk}$ are still homogeneous polynomials of degree $\omega-\de$ and can therefore be written as ${\cal P}_{lk}(S,R)=S^{\omega-\de}{\cal P}_{lk}(1,R/S)$. Since the only remaining unknown function depends on a single variable $R/S$, substituting such functions in (\ref{Schreq}) results in a second order ordinary differential equation (related to the Jacobi polynomial equation) that has a unique polynomial solution, which can be verified to agree (up to normalization) with (\ref{Etarget}). Hence, any function of the form (\ref{cart}) satisfying (\ref{Schreq}) is a target space extension of a linear combination of AdS mode functions of frequency $\omega$. Note that (\ref{Etarget}) is itself manifestly of the form (\ref{cart}).

We could have equivalently rephrased (\ref{cart}) by saying that $\cP(X^2+Y^2,X^i)/(X-iY)^\omega$ has to satisfy the pseudo-Euclidean wave equation on the target space. Whether the Schr\"odinger picture or the wave equation picture is advantageous, depends on the question one is trying to address. The harmonic oscillator Schr\"odinger equations has very explicit symmetries to which we shall return below. In any case, $\cP$ satisfies
\beq
\label{eqf}
4(\sigma\del^2_{\sigma}-(\omega-1)\del_{\sigma})\cP=\sum_i\del^2_i\cP,
\eeq
as well as the homogeneity condition
\beq
\label{eqhm}
(2\s \del_\s+X^i\del_i) \cP=(\om-\de)\cP.
\eeq
For convenience, we have introduced $\sigma\equiv S^2\equiv X^2+Y^2$. Being a polynomial, $\cP$ can be written explicitly as a sum of terms of the form $\s^{(N-\sum p_i)/2} X_1^{p_1}\cdots X_d^{p_d}$, where $N=\om-\de$ and $p_i$ have to be such that $N-\sum p_i$ is non-negative and even. Equation (\ref{eqf}) will then say that the coefficients of the terms with $\sum p_i=N$ are arbitrary, however, once those are specified, the coefficients of the terms with lower values of $\sum p_i$ are completely fixed by (\ref{eqf}). The number of independent solutions is then the same as the number of monomials of the form $X_1^{p_1}\cdots X_d^{p_d}$ with $\sum p_i=N$. This is (\ref{degenr}), which is the total number of independent AdS mode functions of frequency $\omega$, highlighting once again that our representation of the mode functions does not miss anything.

For general values of $\om$, because of the presence of a noninteger power, defining (\ref{cart}) requires a codimension 1 cut in the pseudo-Euclidean embedding space emanating from the (codimension 2) hypersurface $X=Y=0$ and extending to infinity. (Viewed within 2-planes satisfying $X^i=\mbox{const}$ and parametrized by the complex coordinate $X-iY$, the cut looks like an ordinary branch cut from the branching point $X-iY=0$ to infinity. With this cut, $(X-iY)^\omega$ appearing in (\ref{cart}) is single-valued.) Whether this feature demands any special attention depends on the problem one is considering. For example, when dealing with selection rules in the next section, we shall be confined to the case of integer $\om$ (because this is when non-trivial selection rules arise), where the cuts are absent and (\ref{cart}) is well-defined globally.

Solutions to the Schr\"odinger equation (\ref{Schreq}) can be constructed straightforwardly by separation of variables. The relation of such product solutions to the basis (\ref{Etarget}) is highly non-trivial however, and relies on elaborate identities between families of orthogonal polynomials. Since we shall not be using this relation directly, we refer interested readers to appendix B for further details.

The Schr\"odinger equation (\ref{Schreq}) presents the advantage of having a set of symmetries which are easy to characterize. To start with, one has an $SU(d,2)$ group with the generators
\beq
\label{LHSUd2}
H_{IJ}=a^\dagger_Ia_J+a^\dagger_Ja_I
\qquad\mbox{and}\qquad
L_{IJ}=i\left(a^\dagger_Ia_J-a^\dagger_Ja_I\right).
\eeq
Here, $a$ and $a^\dagger$ are creation and annihilation operators defined by
\beq
\label{aXi}
a_I=\frac{1}{\sqrt{2}}\left(\frac{\del}{\del X^I}+\eta_{IJ}X^J\right)
\qquad\mbox{and}\qquad
a_I^{\dagger}=\frac{1}{\sqrt{2}}\left(-\frac{\del}{\del X^I}+\eta_{IJ}X^J\right)
\eeq
and satisfying the commutation relation
\beq
\label{aacom}
[a_I,a^{\dagger}_J]=\eta_{IJ}.
\eeq 
Note that the definition of the creation-annihilation operators is non-standard for the (negative metric  signature) $X$- and $Y$-directions. The $SU(d,2)$ symmetry with the above generators is a straightforward modification of the standard $SU(d+2)$ symmetry for an isotropic oscillator in a $(d+2)$-dimensional Euclidean space.

Since the AdS mode functions (and hence the Higgs oscillator eigenstates) are realized as a subset of solutions of the Schr\"odinger equation (\ref{Schreq}), one should expect that a complete set of symmetries of the mode functions can be recovered as a subset of all symmetries of (\ref{Schreq}). Note however, that we do not mean mere linear combinations of (\ref{LHSUd2}), but rather their arbitrary polynomial combinations, since any product of symmetry generators is also a symmetry generator in the quantum context. We shall give some consideration to this type of nonlinear structures in the last section. Before going in that direction, however, we shall explain how the ordinary $SO(d,2)$ transformations generated by $L_{IJ}$ of (\ref{LHSUd2}), which can also be written more straightforwardly as
\beq
L_{IJ}=i\left(\eta_{IK}X^K\frac{\del}{\del X^J}-\eta_{JK}X^K\frac{\del}{\del X^I}\right),
\label{Lsimp}
\eeq
explain the selection rules in the interaction coefficients (\ref{Ccoeff}).


\section{AdS isometries and selection rules}

Before proceeding with selection rule analysis, we note that a part of the AdS isometry group generated by $L_{IJ}$ of (\ref{Lsimp}) acts as raising and lowering operators on mode functions of the form (\ref{cart}). Indeed, for $\psi$ of the form (\ref{cart}),
acting with $L_{+i}=L_{Xi}+iL_{Yi}$ and $L_{-i}=L_{Xi}-iL_{Yi}$ generates
\begin{eqnarray}
L_{+i}\psi&=&\frac{e^{(\sigma-X^iX^i)/2}}{i(X-iY)^{\omega+1}}\left(\sigma\del_i\cP+2X^i\sigma\del_\sigma \cP-2\omega X^i\cP\right),
\label{raise}
\\
L_{-i}\psi&=&\frac{e^{(\sigma-X^iX^i)/2}}{i(X-iY)^{\omega-1}}\left(\del_i\cP+2X^i\del_\sigma \cP\right).
\label{lower}
\end{eqnarray}
The right-hand sides above are themselves manifestly of the form (\ref{cart}), but with the values of $\om$ shifted by $\pm 1$.
The commutation relation of the raising and lowering operators are given by
\beq
\label{comrl}
[L_{+ i},L_{+ j}]=[L_{- i},L_{- j}]=0,\qquad[L_{+i},L_{-j}]=2(iL_{ij}-\delta_{ij}L_{XY}).
\eeq
$L_{XY}$, which would be called the dilatation operator in the conformal group interpretation of $SO(d,2)$, acts on functions $\psi$ of the form (\ref{cart}) simply as $L_{XY}\psi=\om\psi$.

Using the above algebra of raising and lowering operators, one can construct any mode function of frequency $\de+N$ starting from the lowest mode $\psi_0\equiv e^{(\sigma-X^iX^i)/2}(X-iY)^{-\de}$ of frequency $\de$:
\beq
\psi\sim L_{+i_1}\cdots L_{+i_N}\psi_0.
\label{genert}
\eeq
The mode number counting is manifestly correct, given by the number of totally symmetric rank $N$ tensors. The representation given by (\ref{genert}) is of course directly parallel to the standard construction of descendants from primary operators in conformal field theory, where the $SO(d,2)$ isometries are interpreted as the conformal group. Our formulation in terms of homogeneous polynomials of the embedding space coordinates makes this representation very concrete. (Further systematic considerations of the implications of the conformal group for the AdS dynamics can be found, e.g. in \cite{FKPS}, or the very recent work \cite{HKPS}. A good review of the $SO(d,2)$ representations relevant to the dynamics of fields in AdS can be found in \cite{AdSsup}.)

Armed with the mode function representation of the form (\ref{genert}), we can return to the selection rule problem. We restrict ourselves to the massless field case, where $\de=d$ is an integer, and so are all the frequencies (\ref{omega}). Then, the sum in (\ref{flow}) in principle includes contributions from resonances of the form $\om_{nlk}= \omega_{n_1l_1k_1}+\omega_{n_2l_2k_2}+\cdots+\omega_{n_Nl_Nk_N}$, but these contributions drop out, because the corresponding $C$ coefficients of (\ref{Ccoeff}) in fact vanish.\footnote{Our proof should in fact work equally well for any other value of the mass for which $\de$ is such that there are solutions to the resonance condition with all plus signs. One just needs to replace the number of raising operators needed to produce frequency $\om$ in our derivation below, which is  $\om-d$ for the massless scalar field, by $\om-\de$.} We therefore set out to prove that
\beq
C=\int dx\,d\Omega\, \frac{\tan^{d-1}x}{\cos^2x}\,e_1\cdots e_K
\label{C123}
\eeq
vanishes whenever the frequencies $\om_1,\cdots,\om_K$ corresponding to {\em any} set of $K$ AdS mode functions $e_1,\cdots,e_K$  satisfy $\omega_1=\omega_2+\cdots+\omega_K$. The integral in (\ref{C123}) is inconvenient in that it runs only over one spatial slice of the AdS space (\ref{AdSmetric}). For symmetry-related considerations, it is much more natural to rewrite (\ref{C123}) as an integral over the entire AdS hyperboloid (\ref{embedding}). This is straightforwardly done by defining $\psi_n=e^{i\om_nt}e_n$ and introducing a (trivial) integral over $t$ from $0$ to $2\pi$, resulting in the following representation for $C$ as an integral over the entire AdS hyperboloid (for notation convenience, we shall work up to the overall normalization of $C$ for the rest of this section):
\beq
C=\int dt\,dx\,d\Omega\, \frac{\tan^{d-1}x}{\cos^2x}\,\psi^*_1\psi_2\cdots \psi_K.
\label{Chyper}
\eeq
The integrand is, of course, $t$-independent by virtue of $\omega_1=\omega_2+\cdots+\omega_K$.

We have seen in the previous section that the qualitative properties of the AdS mode functions $\psi_n$ are particularly transparent in terms of Cartesian coordinates in the flat embedding space. We shall now transform (\ref{Chyper}) to a coordinate system which makes the embedding space properties manifest. One welcome circumstance is that, after we have reintroduced $t$ in (\ref{Chyper}), $-\tan^{2(d-1)}x\sec^4x$ is simply the determinant of the metric (\ref{AdSmetric}) of the spacetime over which one is integrating, and $dt\,dx\,d\Omega\tan^{d-1}x\sec^2x$ is simply the corresponding invariant measure that can be easily transformed to other coordinates.

To make use of the simplifications afforded by the flat embedding space, we parametrize the AdS$_{d+1}$ hyperboloid (\ref{embedding}) by the $d$ embedding coordinates $X^i$ and $t$, with the two remaining embedding coordinates given by
\beq
\label{Xyt}
X=\sqrt{1+X^kX^k}\cos t,\qquad
Y=\sqrt{1+X^kX^k}\sin t.
\eeq
The AdS metric can now be written as
\beq
\label{txmetric}
ds^2=-(1+X^kX^k)dt^2+\left(\delta_{ij}-\frac{X^iX^j}{1+X^kX^k}\right)dX^idX^j.
\eeq
Interestingly, if we recall that the AdS spacetime is conformal to a direct product of time and a spatial half-sphere (often referred to as `one half of the Einstein static universe'), then parametrizing the AdS using the above coordinates implies parametrizing the said half-sphere via the so-called {\em gnomonic} projection (casting images from the center of the sphere on a tangent plane). It has been known since \cite{Higgs} that such gnomonic coordinates allow simple expressions for the conserved quantities of the Higgs oscillator. In that context, they emerge as peculiar, even if useful, entities. In our context, the gnomonic coordinates are imposed on us by the geometry under consideration as the simplest Cartesian coordinates in the flat embedding space.

The raising and lowering operators $L_{+i}=L_{Xi}+iL_{Yi}$ and $L_{-i}=L_{Xi}-iL_{Yi}$ can be written through the coordinates (\ref{txmetric}) as
\beq
\label{Lxt}
L_{\pm n}=e^{\pm it}\left(\pm\frac{X^n}{ \sqrt{1+X^kX^k}}\frac{\del}{\del t}-i\sqrt{1+X^kX^k}\frac{\del}{\del X^n}\right).
\eeq
The determinant of the metric (\ref{txmetric}) is simply $-1$, and hence (\ref{Chyper}) can be expressed as\footnote{That the determinant of (\ref{txmetric}) is $-1$ can be quickly seen by noticing that at each point $(t,X^i)$, (\ref{txmetric}) possesses the following eigenvectors: $(1,0)$ with the eigenvalue $-(1+X^kX^k)$, $(0,X^i)$ with the eigenvalue $1/(1+X^kX^k)$ and $(d-1)$ eigenvectors with zero $t$-components orthogonal to $(0,X^i)$, all of which have eigenvalues $1$. The product of all these eigenvalues of (\ref{txmetric}) is simply $-1$.}
\beq
\label{intmode1}
C=\int dt\,dX^i\, \psi^*_1\psi_2\cdots \psi_K.
\eeq
It can be straightforwardly verified that $L_{+i}$ and $L_{-i}$ are Hermitian conjugate with respect to the integration measure in (\ref{intmode1}). Furthermore, being first order differential operators, they possess the obvious property 
\beq
L_{\pm i}(fg)=(L_{\pm i}f)g+f(L_{\pm i}g).
\label{distrL}
\eeq

With the structures we have displayed, one can immediately prove the selection rules in (\ref{intmode1}) and hence in (\ref{Ccoeff}). $\psi_1$ can be written as $\omega_1-d=\omega_2+\cdots+\omega_K-d$ raising operators acting on $\psi_0$, as in (\ref{genert}). Hermiticity properties allow us to turn these into the same number of lowering operators acting on $\psi_2\cdots \psi_K$. By (\ref{distrL}), the lowering operators will simply be distributed among the various factors in the product $\psi_2\cdots \psi_K$ in different ways. However, each $\psi_n$ is given by acting on $\psi_0$ by $\om_n-d$ raising operators, whereas $L_{-i}\psi_0$ is 0. Hence there are simply too many lowering operators coming from $\psi_1$ and they will annihilate all the terms, leaving no non-zero contributions in the final result. This completes our proof of the selection rules for (\ref{Ccoeff}).

We remark that the structure we have displayed above may superficially appear overly constraining. The resonance condition $\omega_1=\omega_2+\cdots+\omega_K$ is minimal in the sense that decreasing $\omega_1$ by just one, i.e., $\omega_1=\omega_2+\cdots+\omega_K-1$, no longer results in selection rules (which can be verified by simple Jacobi polynomial manipulations). However, the number of the lowering operators coming from $\psi_1$ in our above derivation is much greater than what would have been minimally necessary to annihilate the product $\psi_2\cdots \psi_K$, which may seem paradoxical. The resolution is that, if we had imposed $\omega_1=\omega_2+\cdots+\omega_K-1$ in (\ref{C123}) instead of the correct resonance condition, we would not be able to extend the integrals to run over the entire AdS hyperboloid, as in (\ref{Chyper}), and the rest of our derivation would no longer apply.

We comment on the apparent minor discrepancy between the selection rules we have derived and the ones reported on \cite{Yang}. In \cite{Yang}, the condition that $Kd$ (in our present notation) should be even was imposed (in addition to the resonant relation between the frequencies) in order for the selection rules to hold, whereas we see no need for such a condition. The reason, we believe, is that \cite{Yang} has focused exclusively on the radial part of the integral (\ref{C123}) and the corresponding radial parts of the mode functions (\ref{modefunctions}). It is then claimed that the radial ($x$-)integral in (\ref{C123}) vanishes only if $Kd$ is even. We recall that $\om_k=d+2n_k+l_k$, where $n_k$ and $l_k$ are the radial and angular momentum `quantum numbers' of the mode functions (\ref{modefunctions}). If $Kd$ is odd and $\om_1=\om_2+\cdots+\om_K$, it is easy to see that the sum $l_1+l_2+\cdots+l_K$ is odd. However, spherical harmonics of angular momentum $l$ have inversion parity $(-1)^l$. Hence, with all the above specifications, the integrand in (\ref{C123}) has a negative inversion parity and is guaranteed to vanish upon the integration over the angles. The constraint on $Kd$ is therefore unnecessary for formulating the selection rules (and only the resonance condition $\om_1=\om_2+\cdots+\om_K$ has to be satisfied). Our derivations have taken advantage of the embedding space representation for the mode functions that does not require separating the mode functions into radial and angular parts, and the selection rules that emerge automatically take into account all possible constraints from the integrations in (\ref{C123}).


\section{Isometries, hidden symmetries and the Higgs oscillator algebra}

Having established the selection rules, we would like to return to the question of algebraic structures and hidden symmetries of the Higgs oscillator, and see in particular what light can be shed on these questions using the connections to the AdS spacetime and the flat embedding space we have described above.

The Higgs oscillator is a particle on a $d$-sphere with the polar angle $\theta$ moving in a potential proportional to $1/\cos^2\theta$. It is straightforward to show that all the orbits of this motion are closed and the system possesses a large set of classical conserved quantities, which form an $su(d)$ Lie algebra under taking Poisson brackets \cite{Higgs}. The system is in fact maximally superintegrable, which explains why its trajectories are only dense in 1-dimensional manifolds in the phase space. (General review of superintegrability can be found in \cite{superint}.) Quantum-mechanically, one can solve exactly for the spectrum and discover that the degeneracies of the energy levels are again dictated by a hidden $SU(d)$ symmetry, however, no explicit closed form construction for the corresponding symmetry generators is known \cite{Higgs, Leemon, interbasis}.

It was demonstrated in \cite{EK} that the mode functions of a scalar field in Anti-de Sitter spacetime are in a one-to-one correspondence with the Higgs oscillator energy eigenstates. The relation can be read off (\ref{schrod}-\ref{hggs}). In particular, the multiplet structures are completely identical. The AdS$_{d+1}$ spacetime enjoys many special properties such as an $SO(d,2)$ isometry group and a simple embedding as a hyperboloid in a $(d+2)$-dimensional flat space. We have extended the AdS mode functions in the embedding space and found a very simple structure in terms of homogeneous polynomials (\ref{cart}). The extended functions satisfy a harmonic oscillator Schr\"odinger equation (with a pseudo-Euclidean signature) on the embedding space, given by (\ref{Schreq}). The AdS isometries provide a spectrum-generating algebra for the Higgs oscillator, with raising and lowering operators given by (\ref{raise}-\ref{lower}). (Discussion of spectrum-generating algebras, including generalized orthogonal algebras, can be found in \cite{spectrum}.)

It would be natural to attempt characterizing the symmetries of the Higgs oscillator as a subset of symmetries of the Sch\"odinger equation (\ref{Schreq}). Indeed, since every Higgs oscillator energy eigenstate can be related to a certain solution of (\ref{Schreq}), the symmetries of the Higgs oscillator must form a subset of the symmetries of (\ref{Schreq}). One can immediately construct an $SU(d,2)$ group of symmetries of (\ref{Schreq}) with the generators given by (\ref{LHSUd2}). One then should ask which of the generators preserve the functional form of (\ref{cart}), including the homogeneity of the polynomial $\cP$, which is a necessary condition for the corresponding function to represent a Higgs oscillator state. It turns out, however, that only the obvious $SO(d)$ generators $L_{ij}$ of (\ref{Schreq}) preserve the form of (\ref{cart}). Where can one find the remaining $SU(d)$ symmetry generators?

We note that given two symmetry generators for the Schr\"odinger equations (\ref{Schreq}), their ordinary product is also a symmetry generator. We should not therefore limit our search for the Higgs oscillator symmetries to linear combinations of (\ref{LHSUd2}). In particular, looking at the quadratic combinations, one discovers that 
\beq
\Lambda_{ij}=\frac{1}{2}\big(\{L_{Xi},L_{Xj}\}+\{L_{Yi},L_{Yj}\}\big)
\label{Lam}
\eeq
preserves the functional form of (\ref{cart}) and therefore provides a symmetry generator for the Higgs oscillator. Specifically, acting with $\Lambda_{ij}$ on functions of the form (\ref{cart}) returns
\beq
\label{Lambdaact}
\frac{e^{\frac{1}{2}(\sigma-X^iX^i)}}{(X-iY)^{\omega}}\left(\sigma\del_i\del_j\cP+X_iX_j\del_k\del_k \cP+(2\sigma\del_\sigma-\omega+1)(X_i\del_j\cP+X_j\del_i\cP)+\delta_{ij}(2\sigma\del_\sigma-\omega)\cP\right).
\eeq
The commutator of $\Lambda_{ij}$ is given by
\begin{eqnarray}
\label{comlambda}
\dsty [\Lambda_{ij},\Lambda_{kl}]=&i\left(\{\Lambda_{ik},L_{jl}\}+\{\Lambda_{il},L_{jk}\}+\{\Lambda_{jk},L_{il}\}+\{\Lambda_{jl},L_{ik}\}\right)\vspace{2mm}\nonumber\\ \dsty&-i(1+L_{XY}^2)\left(\delta_{ik}L_{jl}+\delta_{il}L_{jk}+\delta_{jk}L_{il}+\delta_{jl}L_{ik}\right).
\end{eqnarray}
While deriving the above formula, one has to use the identity
\beq
L_{Xn}L_{Ym}-L_{Yn}L_{Xm}=-L_{XY}(L_{mn}+i\de_{mn}).
\eeq
The generators $\Lambda_{ij}$  thus do not form a Lie algebra, but rather a quadratic algebra. They are in fact simply the conserved quantities quadratic in momenta considered already in \cite{Higgs}. Here, we have managed to relate these conserved quantities to the spectrum-generating algebra originating from the $SO(d,2)$ isometries of the AdS space. Quadratic algebras have been extensively discussed in the literature in relation to the Higgs oscillator and other superintegrable systems, see, e.g., \cite{quadr,quadr1,quadr2,quadr3,quadr4, quadr5, quadrruan, quadr6,quadr7,quadr8,quadr9}. (A curious application of these algebras to constructing deformations of fuzzy sphere solutions in matrix theories can be found in \cite{fuzzy}.) Using (\ref{eqf}-\ref{eqhm}), one can show that the trace $\Lambda_{ii}$ is not in fact independent, if acting on functions of the form (\ref{cart}), but can be expressed through $L_{ij}L_{ij}$, $L_{XY}$ and $L_{XY}^2$. This agrees with the observations about conserved quantities in \cite{Higgs}. The situation we encounter here is in a way complementary to the more familiar ordinary quantum Coulomb problem, where the standard conserved quantities satisfy a Lie algebra (if one wants to treat the discrete and continuous spectrum on the same footing, one needs to work with loop algebras \cite{loop}), whereas exotic conserved quantities satisfying quadratic Yangian algebras can be constructed \cite{yangian1,yangian2}. For the Higgs oscillator, straightforward quantization of classical conserved quantities leads to a quadratic algebra, while the construction of conserved quantities satisfying an $su(d)$ Lie algebra remains an outstanding problem.

We have generally been writing our generators either in terms of their action on the embedding space functions of the form (\ref{cart}), or in terms of their action on the AdS mode functions $e(x,\Omega)$ extended to the AdS hyperboloid by multiplying with $e^{i\omega t}$, where $\omega$ is the corresponding frequency. Either set of functions is in a one-to-one correspondence with the Higgs oscillator energy eigenstates at energy level number $\omega-\delta$, so in principle our representations are all one needs. However, one can also write a (slightly more awkward) representation for the generators as explicit operators acting on the Higgs oscillator wave functions (with all of the AdS scaffolding completely purged), as we shall briefly explain below.

The action of the raising and lowering operators (\ref{Lxt}) on functions of the form $e^{i\om t}e_{\om}(X^i)$, where $e_{\om}(X^i)$ is any AdS mode function of frequency $\om$ expressed through gnomonic coordinates (\ref{txmetric}), is given by
\beq
\label{Lxtmode}
L_{\pm n}(e^{i\om t}e_{\om}(X^i))=e^{\pm i(\om\pm1)t}\left(\pm\frac{i\om X^n}{\sqrt{1+X^kX^k}}-i\sqrt{1+X^kX^k}\frac{\del}{\del X^n}\right)e_{\om}(X^i).
\eeq
This representation can be used to consistently strip off the $e^{i\om t}$ factors and define the action of $L_{\pm n}$ on $e_{\om}(X^i)$ itself.
We then recall that the relation between the AdS mode functions and the Higgs oscillator wave functions is given by multiplication with $\cos^{(d-1)/2}x\equiv(1+X^kX^k)^{-(d-1)/4}$, as indicated above (\ref{schrod}). So if converting from the AdS language to the Higgs oscillator language, the action of any operator has to be conjugated by this multiplication. Applied to the operator above, this procedure gives
\beq
\label{LHiggs}
L^{\mbox{\scriptsize(Higgs)}}_{\pm n}=\pm\frac{iX^n}{ \sqrt{1+X^kX^k}}\,\hat\om-i(1+X^kX^k)^{(d+1)/4}\frac{\del}{\del X^n}(1+X^kX^k)^{-(d-1)/4}.
\eeq
What remains is to define the operator $\hat\om$ that acts by multiplying the Higgs oscillator energy eigenfunctions by $\de+N$, where $N$ is the energy level number. For the AdS mode functions, this operator was simply $L_{XY}$. For the Higgs oscillator, this operator can be expressed as a function of the Higgs oscillator Hamiltonian. (In practice, energies of the Higgs oscillator are quadratic functions of the energy level number, so the Higgs Hamiltonian is a quadratic function of $\hat\om$, and expressing $\hat\om$ through the Higgs Hamiltonian will involve some square roots.) $L^{\mbox{\scriptsize(Higgs)}}_{+n}$ defined above is Hermitian conjugate to $L^{\mbox{\scriptsize(Higgs)}}_{-n}$ with respect to the standard integration measure on the sphere in gnomonic coordinates ($dX^i(1+X^kX^k)^{-(d+1)/2}$).


\section{Acknowledgments}

We would like to thank Ben Craps, Puttarak Jai-akson, Chethan Krishnan and Joris Vanhoof for collaboration on related subjects. We have benefitted from discussions with Piotr Bizo\'n, Martin Cederwall, George Chechin, Axel Kleinschmidt, Luis Lehner, Maciej Maliborski, Gautam Mandal, Shiraz Minwalla, Karapet Mkrtchyan, Andrzej Rostworowski, and especially Takuya Okuda and Alessandro Torrielli.
The work of O.E. is funded under CUniverse
research promotion project by Chulalongkorn University (grant reference CUAASC). R.N. is supported by a scholarship under the Development and Promotion of Science and
Technology Talents Project (DPST) by the government of Thailand.


\appendix

\section{Geodesics in AdS and the classical Higgs oscillator}

We would like to demonstrate the relation between the geodesic motion in AdS and the orbits of the classical Higgs oscillator.   In both problems, it is convenient to use the rotation symmetry and place the orbit in the equatorial plane, setting the azimuthal angles to $\pi/2$. The motion is then described by the polar angle $\phi$, the `radial' angle called $x$ in the AdS metric (\ref{AdSmetric}) and $\theta$ for the Higgs oscillator, and the time variable.

The geodesic equation in AdS can be obtained by considering the Lagrangian.
\beq
\label{LAdS}
L_{\mbox{\scriptsize{AdS}}}=\frac{1}{\cos^2x}\left(-\left(\frac{dt}{d\lambda}\right)^2+\left(\frac{dx}{d\lambda}\right)^2+\sin^2x\left(\frac{d\phi}{d\lambda}\right)^2\right).
\eeq
where $\lambda$ is the affine parameter. The equations are given by
\beq
\label{geoAdS}
\frac{d^2x}{d\lambda^2}+\tan^2x\left(\frac{dx}{d\lambda}\right)^2+E^2\frac{\sin x}{\cos^3x}-\frac{l^2}{\tan^3x}=0.
\eeq
where $E$ and $l$ are the conserved quantities defined by
\beq
\label{EL}
E=\frac{1}{\cos^2x}\frac{dt}{d\lambda}\qquad\mbox{and}\qquad
l=\tan^2x\frac{d\phi}{d\lambda}.
\eeq
Introducing a new curve parameter $d\tau=\cos x\,d\lambda$, (\ref{geoAdS}) is transformed to
\beq
\label{geoAdS1}
\frac{d^2x}{d\tau^2}+E^2\sin x\cos x-l^2\frac{\cos x}{\sin^3x}=0.
\eeq
The Higgs oscillator equation of motion can be derived from the Lagrangian
\beq
\label{LHiggs}
L_{\mbox{\scriptsize{Higgs}}}=\frac{1}{2}\left(\left(\frac{d\theta}{dt}\right)^2+\sin^2\theta\left(\frac{d\phi}{dt}\right)^2\right)-\frac{k}{\cos^2\theta}.
\eeq
The equation of motion is given by
\beq
\label{Higgseqn}
\frac{d^2\theta}{dt^2}-l^2\frac{\cos\theta}{\sin^3\theta}+\frac{2k\sin\theta}{\cos^3\theta}=0,
\eeq
where $l=(d\phi/dt)\sin^2\theta$. Integrating (\ref{geoAdS1}) and (\ref{Higgseqn}) with respect to $\tau$ and $t$, respectively, yields
\begin{eqnarray}
\label{geoAdS2}
\frac{1}{2}\left(\frac{dx}{d\tau}\right)^2-\frac{E^2}{2}\cos^2x+\frac{l^2}{2\sin^2x}&=&C_{\mbox{\scriptsize{AdS}}},\\
\label{Higgseqn2}
\frac{1}{2}\left(\frac{d\theta}{dt}\right)^2+\frac{l^2}{2\sin^2\theta}+\frac{k}{\cos^2\theta}&=&C_{\mbox{\scriptsize{Higgs}}}.
\end{eqnarray}
Integrating (\ref{geoAdS3}) and (\ref{Higgseqn2}) further and rewriting the result as an equation for the orbital shapes $x(\phi)$ and $\theta(\phi)$, one obtains
\begin{eqnarray}
\label{geoAdS3}
\phi_{\mbox{\scriptsize{AdS}}}&=&\int\frac{l\,dx}{\sin^2x\sqrt{E^2-l^2\mbox{cosec}^2x+(2C_{\mbox{\scriptsize{AdS}}}-l^2)\sec^2x}},\\
\label{Higgseqn3}
\phi_{\mbox{\scriptsize{Higgs}}}&=&\int\frac{l\,d\theta}{\sin^2\theta\sqrt{2C_{\mbox{\scriptsize{Higgs}}}-l^2\mbox{cosec}^2\theta-2k\sec^2\theta}}.
\end{eqnarray}
These equations are the same, given an appropriate identification of the integration constants and the strength of the Higgs oscillator potential.

We note that, while the orbital shapes are the same, the AdS geodesic equation generates, in a sense, a superior time evolution. Not only are all orbits in AdS closed, but their periods with respect to $t$ are also the same, unlike the time evolution generated by the Higgs oscillator Hamiltonian. Correspondingly, in the quantum theory, the Higgs oscillator has a spectrum with unevenly spaced levels. The mode functions of the Klein-Gordon equation in the AdS space, which are directly related to the Higgs oscillator energy eigenstates, possess a spectrum of equidistant frequency levels.

We also note in passing a curious self-similarity transformation afforded by the orbits of the Higgs oscillator: under the transformation $\tilde\theta=\arctan(\sin\theta)$, Higgs oscillator orbits get mapped to Higgs oscillator orbits (while the integration constants and the strength of the Higgs oscillator potential get redefined). However, the new variable $\tilde\theta$ only ranges over angles less than $\pi/4$. Thus every orbit of the original system living on a half-sphere ($\theta<\pi/2$) can be recovered from an orbit moving entirely within the cone $\theta<\pi/4$, with an appropriate re-identification of parameters. This self-similarity transformation can, of course, be repeated, yielding motion in even smaller domains containing all the information about the orbits of the original theory.

\section{AdS$_{d+1}$ mode functions in terms of $(d+2)$-dimensional harmonic oscillator}

We would like to ask how to isolate (\ref{Etarget}) among the solutions of (\ref{Schreq}), of which they form a subset.
One can start by inspecting product solutions $\Phi=\xi(S,T)\psi(R,\Omega)$ with
\begin{eqnarray}
\label{eqST}
&\dsty\Big\{-\frac{1}{S}\frac{\partial}{\partial S}\left(S\frac{\partial}{\partial S}\right)-\frac{1}{S^2}\frac{\partial^2}{\partial t^2}+S^2\Big\}\xi=\left((2\delta-d-2)+2\zeta\right)\xi,\\
\label{eqRO}
&\dsty\Big\{\frac{1}{R^{d-1}}\frac{\partial}{\partial R}\left(R^{d-1}\frac{\partial}{\partial R}\right)+\frac{1}{R^2}\Delta_{\Omega_{d-1}}-R^2\Big\}\psi
=-2\zeta\psi.
\end{eqnarray}
Upon examining (\ref{Etarget}), for $\psi$ one can take a perfectly standard isotropic harmonic oscillator basis involving the associated Laguerre polynomials:
\beq
\label{solRO}
\psi_{nlm}(R,\Omega)=R^le^{-R^2/2}L^{(l+\frac{d}{2}-1)}_n(R^2)Y_{lm}(\Omega),\qquad
\zeta=\frac{d}{2}+2n+l.
\eeq
A similar approach to $\xi$ would have failed, however, since it can give neither the $1/S^{\om}$ pole, nor the growing exponential $e^{S^2/2}$ in (\ref{Etarget}). Instead, one should be looking for special solutions of the form
\beq
\label{defchi}
\xi=\frac{e^{i\omega T}}{S^{\omega}}e^{S^2/2}\chi(S),
\eeq
where $\chi$ is a polynomial. Similar special singular solutions growing in classically forbidden regions have been previously considered as a quantum-mechanical curiosity in \cite{continue1,continue2, maseno}. (\ref{eqST}) then becomes
\beq
\label{eqc}
\frac{d^2\chi}{dS^2}+\left(2S-\omega+\frac{1}{2}\right)\frac{d\chi}{dS}-2(\omega-\delta-2n-l)\chi=0.
\eeq
It is easy to verify that polynomial solutions to this equation will only exist if $\omega$ satisfies the quantization condition $\omega=\delta+2j+l$ for some integer $j$.
Rewriting the above equation in terms of $\xi=-S^2$ produces the associated Laguerre polynomial equation
\beq
\label{eqLaguerre}
\xi\frac{d^2\chi}{d\xi^2}+(-\de-2j-l+1-\xi)\frac{d\chi}{d\xi}+({j-n})\chi=0
\eeq
with solutions
\beq
\label{chi}
\chi_{n}=L^{(-\delta-2j-l)}_{j-n}(-S^2).
\eeq

One can then ask how to construct the extended mode functions (\ref{Etarget}) from our special set of factorized solutions built from (\ref{solRO}), (\ref{defchi}) and (\ref{chi}). To this end, we first point out a relation between the Laguerre and Jacobi polynomials, which is a slight generalization of the one that had appeared in \cite{laguerre}.
Consider the generating functions of Jacobi polynomials and Laguerre polynomials,
\beq
\label{Jgen}
\sum^{\infty}_{j=0}\frac{\dsty \big((y-x)t\big)^j P^{(\alpha,\beta)}_j\left(\frac{y+x}{y-x}\right)}{\dsty \Gamma(j+\alpha+1)\Gamma(j+\beta+1)}=\frac{\dsty I_\beta(2\sqrt{yt})I_\alpha(2\sqrt{xt})}{\dsty \sqrt{t^{\alpha+\beta}x^\alpha y^\beta}}
\eeq
and
\beq
\label{Lgen}
\sum^{\infty}_{j=0}L^{(\alpha)}_j(x)\frac{t^j}{\Gamma(j+\alpha+1)}=\frac{e^t}{(xt)^{\alpha/2}}J_\alpha(2\sqrt{xt}).
\eeq
Using the Bessel function relation $J_\alpha(ix)=i^{\alpha}I_\alpha(x)$ and combining (\ref{Jgen}) and (\ref{Lgen}) yields
\beq
\label{JLiden}
P^{(\alpha,\beta)}_j\left(\frac{y+x}{y-x}\right)=\frac{1}{(y-x)^j}\sum^{j}_{n=0}(n+\alpha+1)_{j-n}(j-n+\beta+1)_{n}(-1)^nL^{(\alpha)}_n(x)L^{(\beta)}_{j-n}(-y)
\eeq
where $(x)_n\equiv{\Gamma(x+n)}/{\Gamma(x)}=x(x+1)\cdots(x+n-1)$ is the Pochhammer symbol.

Constructing a quadratic convolution of (\ref{solRO}), (\ref{defchi}) using the above relation, with $x$ identified with $R^2$ and $y$ identified with $S^2$ thus indeed produces a Jacobi polynomial, as in (\ref{Etarget}), but this polynomial (superficially) has strange negative order-dependent weight $(l+\frac{d}{2}-1,-\de-2j-l)$ and a `wrong' argument. To connect it explicitly to (\ref{Etarget}), one needs to further apply the following relation:
\beq
\label{transformedJ}
P^{(\alpha,\beta)}_j\left(1-2\eta\right)=\left(1-\eta\right)^jP^{(\alpha,\alpha-\beta-2j-1)}_j\left(\frac{1+\eta}{1-\eta}\right).
\eeq
This relation is proved by first noticing that the right-hand side is a polynomial of degree $j$ and then verifying that it satisfies the Jacobi equation for weights $(\alpha,\beta)$ with respect to the argument on the left-hand side. This establishes the identity above up to normalization, which can be verified to be correct by comparing the coefficients of one particular power of $\eta$ on the two sides.
Combining (\ref{JLiden}) and (\ref{transformedJ}), one can conclude that
\beq
\label{JLiden2}
S^{2j}P^{(\alpha,-\alpha-\beta-2j-1)}_j\left(1-\frac{2R^2}{S^2}\right)=\sum^{j}_{n=0}(n+\alpha+1)_{j-n}(j-n+\beta+1)_{n}(-1)^nL^{(\alpha)}_n(R^2)L^{(\beta)}_{j-n}(-S^2).\nonumber
\eeq
Note that while this identity has been derived by a succession of rather complicated and not entirely explicit steps, the result can be straightforwardly verified, and holds, by evaluating both sides for special values of the indices using a computer algebra system.
Putting everything together, each extended mode functions (\ref{Etarget}) can be expressed explicitly through solutions of the Schr\"odinger equation (\ref{Schreq}) as
\beq
{\cal E}_{jlk}=\frac{e^{i\omega T}}{S^{\omega}}e^{(S^2-R^2)/2}\sum^{j}_{n=0}W_{nj}L^{(-\delta-2j-l)}_{j-n}(-S^2)L^{(l+\frac{d}{2}-1)}_{n}(R^2)R^lY_{lk}(\Omega).
\label{modeext}
\eeq
where $W_{nj}=(n-\delta-2j-l+1)_{j-n}(j-n+l+\frac{d}{2})_{n}(-1)^n$ (and we have not kept track of the overall normalization).

\end{document}